\documentclass[notitlepage, 
noruninaddress,
bibnotes,
prb,5pt, 
twocolumn,
showpacs,preprintnumbers,superscriptaddress,
amsmath,amssymb,floatfix]{revtex4-1}
\usepackage{amsmath}
\usepackage{graphicx}
\usepackage{ dsfont }
\usepackage{dcolumn}
\usepackage{bm}
\usepackage{color,soul}
\usepackage{xcolor}
\usepackage{url}
\usepackage{tabularx}
\usepackage{array}
\usepackage{booktabs}
\usepackage{comment}
\usepackage{hyperref}
\usepackage{footnote}
\usepackage{lipsum}
\usepackage[normalem]{ulem}
\definecolor{colorA}{rgb}{0.006, 0.5, 0.006}

\newcounter{firstbib}

\usepackage{titlesec}
\newcommand*{\justifyheading}{\raggedright}
\titleformat{\section}
  {\normalfont\bfseries\justifyheading}{\thesection}{1em}{}  
\usepackage[singlelinecheck=false, justification=RaggedRight, format=plain]{caption}
\DeclareCaptionLabelSeparator{bar}{ $\boldsymbol{\mid}$ }
\newcommand{\thesupplementbibliography}{\thebibliography}
\usepackage{etoolbox}
\makeatletter
\apptocmd{\thesupplementbibliography}{\global\c@NAT@ctr 30\relax}{}{}
\makeatother
\begin{document}

\title{
Magnetic particle-antiparticle creation and annihilation
}

\author{Fengshan~Zheng}
\email{f.zheng@fz-juelich.de}
\affiliation{Ernst Ruska-Centre for Microscopy and Spectroscopy with Electrons and Peter Gr\"unberg Institute, Forschungszentrum J\"ulich, 52425 J\"ulich, Germany}

\author{Nikolai~S.~Kiselev}
\email{n.kiselev@fz-juelich.de}
\affiliation{Peter Gr\"unberg Institute and Institute for Advanced Simulation, Forschungszentrum J\"ulich and JARA, 52425 J\"ulich, Germany}

\author{Luyan~Yang}
\affiliation{Ernst Ruska-Centre for Microscopy and Spectroscopy with Electrons and Peter Gr\"unberg Institute, Forschungszentrum J\"ulich, 52425 J\"ulich, Germany}

\author{Vladyslav~M.~Kuchkin}
\affiliation{Peter Gr\"unberg Institute and Institute for Advanced Simulation, Forschungszentrum J\"ulich and JARA, 52425 J\"ulich, Germany}
 \affiliation{Department of Physics, RWTH Aachen University, 52056 Aachen, Germany}

\author{Filipp~N.~Rybakov}
\affiliation{Department of Physics and Astronomy, Uppsala University, SE-75120 Uppsala, Sweden}
\affiliation{Department of Physics, KTH-Royal Institute of Technology, SE-10691 Stockholm, Sweden}

\author{Stefan~Bl\"ugel}
\affiliation{Peter Gr\"unberg Institute and Institute for Advanced Simulation, Forschungszentrum J\"ulich and JARA, 52425 J\"ulich, Germany}

\author{Rafal~E.~Dunin-Borkowski}
\affiliation{Ernst Ruska-Centre for Microscopy and Spectroscopy with Electrons and Peter Gr\"unberg Institute, Forschungszentrum J\"ulich, 52425 J\"ulich, Germany}

\date{\today}

\maketitle

\textbf{
A fundamental property of particles and antiparticles, such as electrons and positrons, is their ability to annihilate one another. 
Similar behavior is predicted for magnetic solitons~\cite{Kovalev_90}-- localized spin textures that can be distinguished by their topological index $Q$.
Theoretically, magnetic topological solitons with opposite values of $Q$, such as skyrmions~\cite{Bogdanov_89} and their antiparticles -- antiskyrmions -- are expected to be able to merge continuously and to annihilate~\cite{Kuchkin_20i}.
However, experimental verification of such particle-antiparticle pair production and annihilation processes has been lacking.
Here, we report the creation and annihilation of skyrmion-antiskyrmion pairs in an exceptionally thin film of the cubic chiral magnet B20-type FeGe observed using transmission electron microscopy.
Our observations are highly reproducible and are fully consistent with micromagnetic simulations.
Our findings provide a new platform for fundamental studies of particles and antiparticles based on magnetic solids and open new perspectives for practical applications of thin films of isotropic chiral magnets.
}

The stability of magnetic skyrmions in B20-type crystals results from a competition between Heisenberg exchange and chiral Dzyaloshinskii-Moriya interaction (DMI)~\cite{Dzyaloshinskii,Moriya}.
Since the cubic anisotropy in such crystals is typically negligibly small and, to a first approximation, Heisenberg exchange and DMI are assumed to be isotropic, it is common to refer to them as \textit{isotropic} chiral magnets. 
In such systems, DMI is predicted to favor skyrmion solutions of fixed chirality~\cite{Bogdanov_94}, in agreement with experimental observations~\cite{Yu_10,Yu_11,Park_14,Tokura_20}.

Skyrmions in isotropic chiral magnets typically take the form of vortex-like tubes or strings, which penetrate through the entire sample thickness.
As a result of conical modulations, a cross-section of an isolated skyrmion tube resembles a two-dimensional (2D) skyrmion in a tilted ferromagnetic vacuum~\cite{Du_18}.
Recent theoretical studies~\cite{Barton-Singer_20, Kuchkin_20i}  of a 2D model of an isotropic chiral magnet have revealed many intriguing effects.  
In particular, it was shown that there is a stable solution for a skyrmion antiparticle -- an antiskyrmion -- which is characterized by opposite chirality in different spatial directions~\cite{Kuchkin_20i}. 

\begin{figure*}[ht]
    \centering
    \includegraphics[width=18cm]{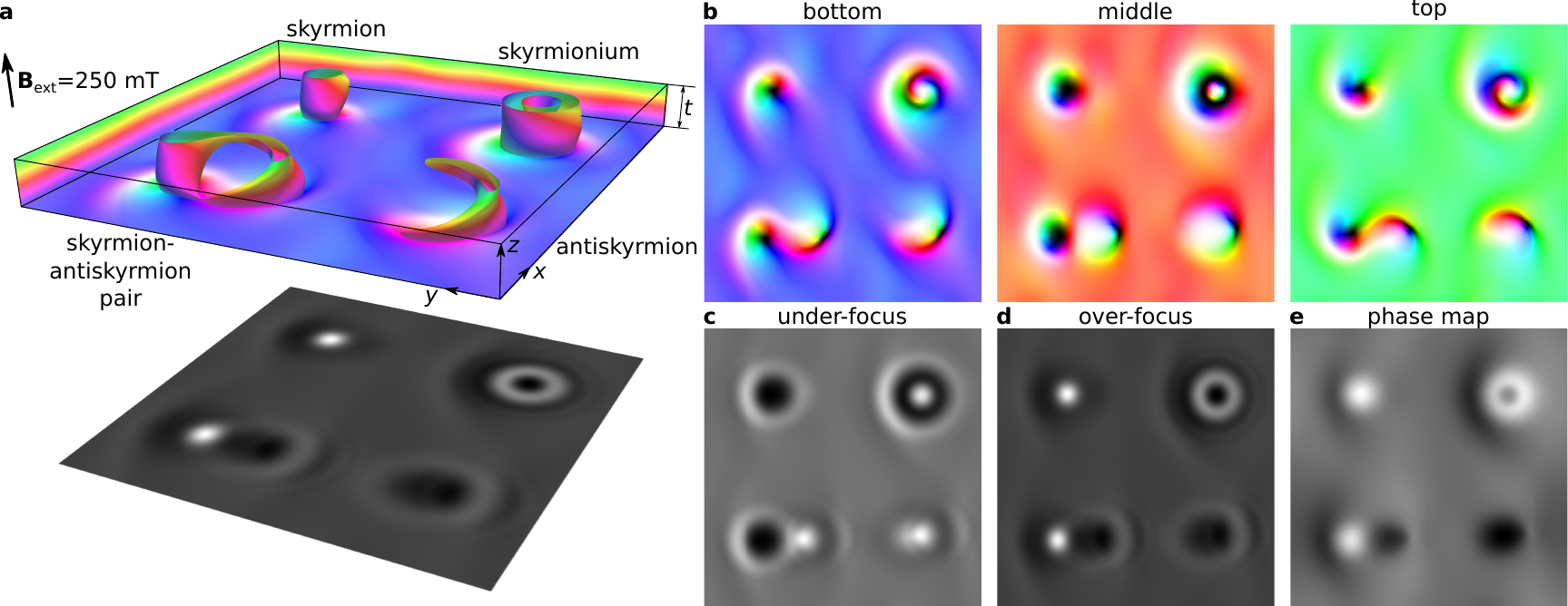}
    \caption{
    \small\textbf{Different localized magnetic states in a thin plate of a chiral magnet.} 
    \textbf{a}, Micromagnetic simulation of a plate of thickness $t$~=~50~nm of an isotropic chiral magnet that supports a skyrmion, an antiskyrmion, a skyrmion-antiskyrmion pair and a skyrmionium in a perpendicular magnetic field. The lateral size of the simulated domain is $512 \times 512$ nm$^2$. Periodic boundary conditions are applied in the $xy$~plane. The spin textures are visualized in the form of isosurfaces $m_\mathrm{z}=0$. The lower image shows an over-focus Lorentz TEM image calculated for the above spin texture for an electron beam that is parallel to the $z$~axis. 
    \textbf{b}, Sections of the spin textures in \textbf{a} at $z=0$ (bottom), $z=t/2$ (middle) and $z=t$ (top). The direction of magnetization is shown using a standard color code: black and white denote up and down spins, respectively, while red-green-blue denote the azimuthal angle.
    \textbf{c}-\textbf{e}, Under-focus and over-focus Lorentz TEM images calculated for defocus distances of 800~$\mu$m and
    and a corresponding calculated electron holographic phase shift image.
    }
    \label{Fig1}
\end{figure*}

We begin by checking the stability of such a solution for a film of finite thickness, taking into account demagnetizing fields. 
Figure~\ref{Fig1}\textbf{a} illustrates statically stable solutions for a skyrmion, an antiskyrmion, a skyrmionium and a skyrmion-antiskyrmion pair obtained by micromagnetic calculations (see Methods and Extended Data Fig.~\ref{FigS-ini}).
(For illustrative purposes, the different spin textures are combined in a single simulated domain in an optimal field at which they are all stable).
The three-dimensional (3D) spin textures are visualized by means of isosurfaces and a standard color code for spin directions~\cite{Zheng_21}.

The color variation at the edges of the simulation indicates the presence of a conical spiral in the direction of the external magnetic field $\mathbf{B}_\mathrm{ext}||\mathbf{e}_\mathrm{z}$.
The period of the cone modulations, $L_\mathrm{D}=4\pi\mathcal{A}/\mathcal{D}$, depends on the ratio between the exchange stiffness constant $\mathcal{A}$ and the DMI constant $\mathcal{D}$. It therefore varies between different compounds. For example, in FeGe $L_\mathrm{D}=70$~nm.
As a result of the presence of conical modulations, demagnetizing fields and a chiral surface twist, the spin texture changes significantly through the film thicknes, as shown in Fig.~\ref{Fig1}\textbf{b}.
Figures~\ref{Fig1}\textbf{c}-\textbf{d} show simulated Lorentz transmission electron microscopy (TEM) images and an electron holographic phase shift image calculated using the approach described in Ref.~\cite{Zheng_21}.
Extended Data Fig.~\ref{FigS-LTEM-defocus} shows a series of Lorentz TEM images simulated for different defocus distances.

Following a general approach for the classification of solutions in systems in which the order parameter is the unit vector field, the localized magnetic textures shown in Fig.~\ref{Fig1} can be classified based on topological index:
\begin{equation}
    Q = \frac{1}{4\pi}\int 
    \mathbf{m}\cdot \left(\partial_x\mathbf{m}\times\partial_y\mathbf{m}\right)  \mathrm{d}x\mathrm{d}y~,
    \label{Q}
\end{equation}
where $\mathbf{m}(\mathbf{r})$ is the magnetization unit vector field. Since the magnetic texture shown in Fig.~\ref{Fig1} is smooth and free of Bloch points, the above definition of $Q$ is valid in any arbitrarily chosen $xy$~plane. 
The total topological index of the combined spin texture shown in Fig.~\ref{Fig1} is zero, since the topological indices of a skyrmion and an antiskyrmion are $-1$ and $+1$, respectively, while the topological index of a skyrmionium is zero.

In our micromagnetic simulations, the skyrmion-antiskyrmion pair always annihilates with increasing magnetic field, while the topological index of the system remains unchanged.
In contrast, the isolated skyrmion and antiskyrmion remain stable over a much wider range of fields.
The behavior of the skyrmion-antiskyrmion pair is consistent with the prediction for the 2D model, suggesting that on a qualitative level a 2D model of a chiral magnet is able to capture the main features of a more advanced 3D model.

Below, we present experimental results on the creation and annihilation of a skyrmion-antiskyrmion pair, which are guided by a different theoretical prediction of a 2D model, which shows that a set of closed domain walls, such as a skyrmion bag, can decay into ``elementary'' particles --  skyrmions and antiskyrmions -- in a certain external magnetic field.
This decay conserves the total topological index and thus represents a homotopic transition.
Extended Data  Figs~\ref{FigS-2D-1} and \ref{FigS-2D-2} illustrate such a decay induced by a tilted magnetic field for a skyrmion bag with $Q=1$ and a skyrmionium with $Q=0$.
We find that, in a sample of finite thickness, this instability occurs even in a perpendicular applied field.
Simulations for a film of finite thickness in the presence of a demagnetizing field are provided in Extended Data Fig.~\ref{FigS-Nucleation-theory}.
Starting with a complex magnetic configuration in zero field, we observe qualitatively the same evolution with applied field as for the 2D case, leading eventually to the formation of skyrmions and antiskyrmions.  

In order to perform experimental observations of the theoretically-predicted phenomena, a thin plate was prepared from a single crystal of B20-type FeGe using a focused ion beam workstation and a lift-out method~\cite{Du_15}.
The nominal thickness of the square plate $t$ is comparable to the size of the chiral modulations $L_\mathrm{D}$ in this compound of 70~nm.
It should be noted that such an exceptionally thin film of a B20-type chiral magnet has not been studied before using Lorentz TEM or other microscopy techniques.
Taking into account possible errors in the thickness estimation of $\sim5$~nm and the likely presence of a thin damaged surface layer of $\sim5$~nm due to sample preparation~\cite{Wolf_21}, it is reasonable to assume that the true magnetic thickness of the FeGe plate is $\sim50$~nm.

\begin{figure*}[ht]
    \centering
    \includegraphics[width=18cm]{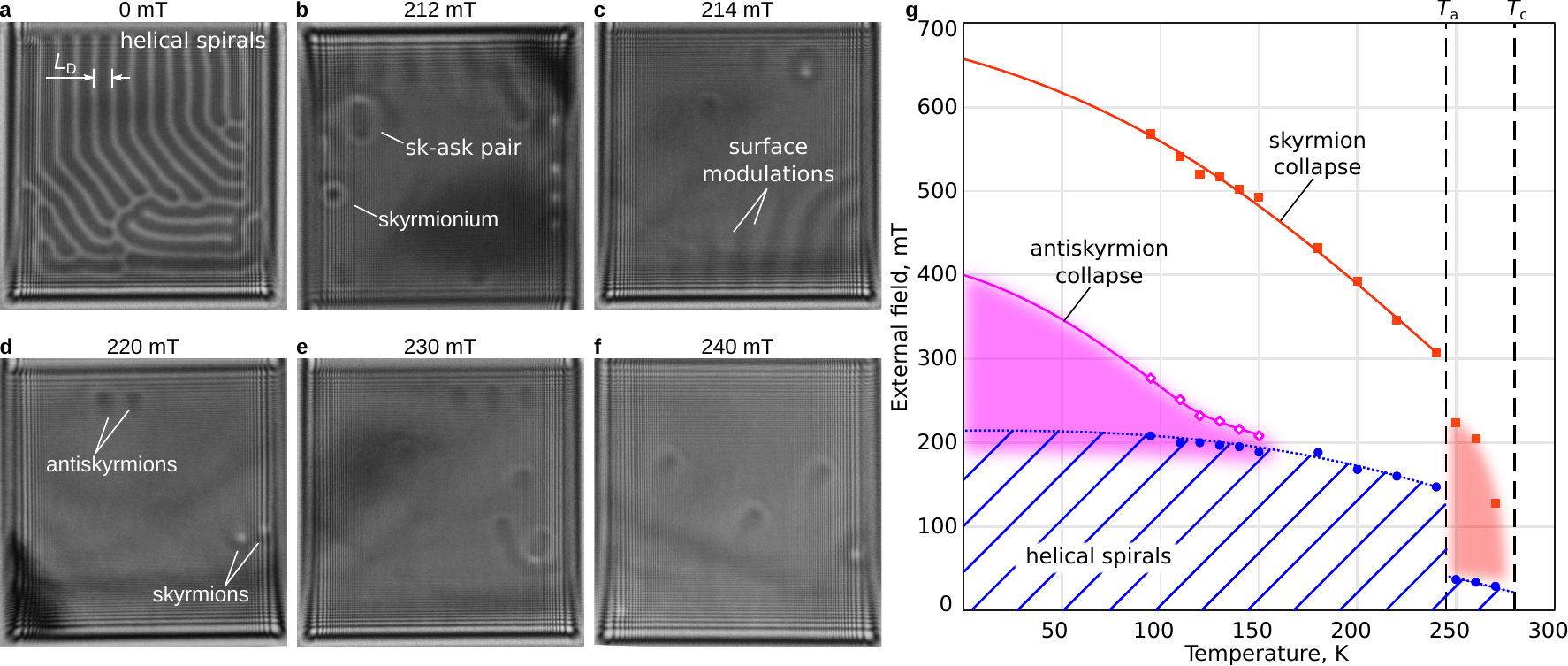}
    \caption{
    \small\textbf{Experimental Lorentz TEM images and diagram of observable states in a thin FeGe sample.}
    \textbf{a}-\textbf{f}, Representative over-focus Lorentz TEM images in a square 1$\mu$m$\times$1$\mu$m-FeGe sample of nominal thickness 70~nm recorded at $T=95$~K. 
    \textbf{g}, Temperature-field diagram of states observed with increasing external magnetic field.
    \textbf{a} is a representative example of a ground state with helical spirals.
    \textbf{b}-\textbf{f} are representative images arranged in order of increasing magnitude of external magnetic field, demonstrating the coexistence of skyrmions and antiskyrmions. 
    \textbf{b} shows a skyrmionium and a skyrmion-antiskyrmion pair. 
    \textbf{c} and \textbf{e} also show skyrmion-antiskyrmion pairs.
    \textbf{d} shows two skyrmions and two antiskyrmions separated by a large distance.
    In \textbf{g}, red and magenta squares indicate the collapse fields of skyrmions and antiskyrmions, respectively.
    The antiskyrmion collapse field is extrapolated towards $T=0$~K based on micromagnetic estimations.
    The dashed region indicates the presence of helical spirals, which transform into surface modulations at high fields, as shown in \textbf{b} and \textbf{c} in the form of weak contrast features whose periodicity is larger than that of the helical spiral $L_\mathrm{D}$.
    The vertical dashed line $T_\mathrm{c}=287$ K marks the Curie temperature of FeGe.
    $T_\mathrm{a}$ is an activation temperature, above which a skyrmion lattice emerges spontaneously in the shaded red region. The symbols correspond to experimentally-measured values, while the lines are guides to the eye.
    }
    \label{FigExp1}
\end{figure*}

\begin{figure*}[ht]
    \centering
    \includegraphics[width=16cm]{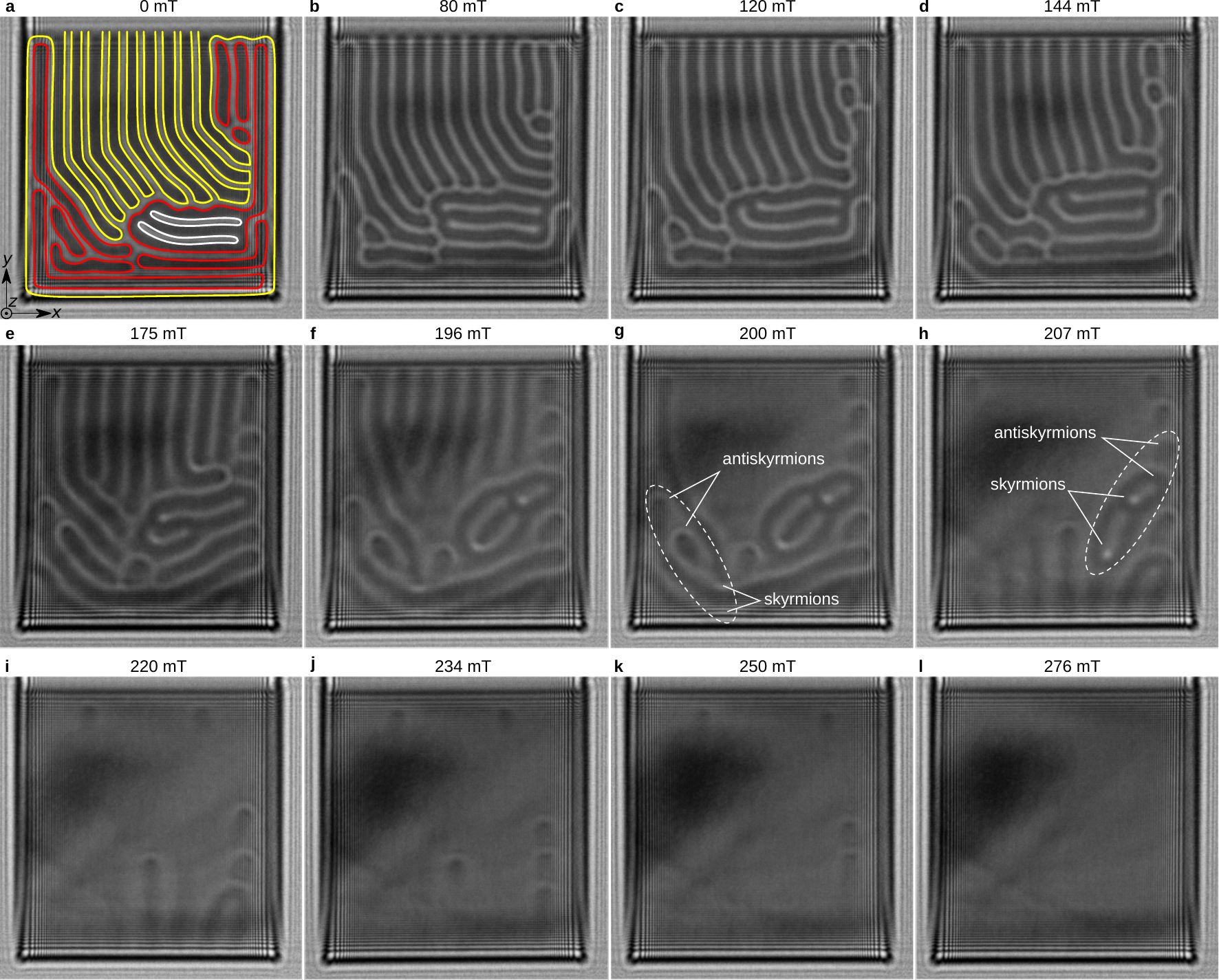}
    \caption{
    \small \textbf{Protocol for antiskyrmion nucleation and skyrmion-antiskyrmion pair production.}
    The figures shows a sequence of over-focus Lorentz TEM images recorded as a function of increasing external magnetic field $\mathbf{B}_\mathrm{ext} || \mathbf{e}_z$.
    The initial configuration in zero field in \textbf{a} is identical to Fig.~\ref{FigExp1}\textbf{a}.
    The white, red and yellow lines mark contours of 180$^\circ$-domain walls.
    The red and white contours form closed loops, while the yellow contours are discontinuous. 
    With increasing field, the white and yellow contours converge to form skyrmions, while the red contours converge to form antiskyrmions.
    After the annihilation of skyrmions and antiskyrmions (\textbf{g} and \textbf{h}), the system converges to six antiskyrmions (\textbf{i} and \textbf{j}). 
    On increasing the field further, the antiskyrmions collapse at $B_\mathrm{ext} > 276$ mT.
    }
    \label{FigExp2}
\end{figure*}

Figure~\ref{FigExp1} show representative experimental over-focus Lorentz TEM images of a square $1~\mu$m $\times$ $1~\mu$m plate of the FeGe sample recorded after distinct cycles of applied external field.
Figure~\ref{FigExp1}\textbf{a} shows a representative ground state of the system in zero field, while \textbf{b}-\textbf{f} show typical contrast in an external magnetic field of above 200~mT applied perpendicular to the plate. Further experimental images are provided in Extended Data Fig.~\ref{FigS-collection}.
Excellent agreement is obtained between theoretical (Fig.~\ref{Fig1}) and experimental (Fig.~\ref{FigExp1}) Lorentz TEM images, confirming the formation of four distinct states. 
In the over-focus regime, an ordinary skyrmion is imaged as a bright circular spot, while an antiskyrmion is imaged as an elongated dark spot with weak bright contrast on only one side.
The magnetic contrast of an antiskyrmion in our isotropic case differs from that of an antiskyrmion in a system with anisotropic DMI~\cite{Nayak_17, Karube_21}, as a result of the asymmetry of antiskyrmions and additional modulations through the film thickness.
The situation is different for Heusler materials~\cite{Nayak_17, Karube_21}, in which antiskyrmions have a fixed orientation to the crystallographic axes. In contrast, in isotropic chiral magnets an antiskyrmion has an additional rotational degree of freedom. Antiskyrmions with different orientations can be seen in Figs~\ref{FigExp1}\textbf{b},~\textbf{c},~\textbf{f}.
A skyrmionium is imaged as a bright circular halo surrounding a dark spot, while a skyrmion-antiskyrmion pair shows a superposition of skyrmion and antiskyrmion contrast.
The TEM images allow a skyrmionium and a skyrmion-antiskyrmion pair to be distinguished despite their topological equivalence.
Extended Data Fig.~\ref{FigS-TEM} shows other representative over- and under-focus Lorentz TEM images, as well as phase shift images recorded using off-axis electron holography.
Extended Data Fig.~\ref{FigS-temperature} shows experimental images of antiskyrmions recorded at different sample temperatures.

Figure~\ref{FigExp1}\textbf{g} shows the experimentally-estimated collapse fields of skyrmions and antiskyrmions as a function of temperature.
The trend line of the experimental points for the antiskyrmion collapse field, extrapolated to zero temperature, shows good agreement with the value of $\sim 400$~mT provided by micromagnetic simulations, as shown in Extended Data Fig.~\ref{FigS-Nucleation-theory}. 
At a magnetic field of $\sim 200$~mT (blue circles), weak contrast associated with surface-induced modulations~\cite{Rybakov_16, Kovacs_review,Turnbull_21} emerges. This contrast transforms into helical spirals when the field is reduced, making it difficult to estimate the lower bound field for antiskyrmion stability, which is indicated by a blurred lower edge of the magenta region.

The experimental images shown in Fig.~\ref{FigExp1} and Extended Data Figs~\ref{FigS-collection}-\ref{FigS-TEM} were obtained by using the following approach, which allows the observed magnetic states to be generated reproducibly.
First, a field of $B_\mathrm{ext}$ $\sim$~50~mT was cycled several times until a pattern of closed domain walls was observed, similar to that shown in Fig.~\ref{FigExp2}\textbf{a}, in which contour lines marked in white, red and yellow follow 180$^\circ$-domain walls.
As a result of the presence of Fresnel fringes from the sample edges, the yellow contours near the left, right and lower edges of the sample only become evident with increasing field, as shown in Figs~\ref{FigExp2}\textbf{b}-\textbf{e}.  
The white and yellow contours enclose areas in which the magnetization is opposite to $\mathbf{B}_\mathrm{ext}$, while the red contours enclose areas in which the magnetization is along $\mathbf{B}_\mathrm{ext}$. In Fig.~\ref{FigExp2}, the external magnetic field points towards the reader.

When the external magnetic field is increased, the \textit{white} contours converge to form skyrmions, while the \textit{red} contours converge to form antiskyrmions.
The larger the number of red contours is present in the initial state, the more antiskyrmions are observed with increasing field and \emph{vice versa}. 
Extended Data Figs~ \ref{FigS-nucleation-1}-\ref{FigS-nucleation-3} illustrate antiskyrmion nucleation for different initial states.
If the helical modulations in the initial state do not form closed loops, then either only skyrmions or no skyrmions at all are observed with increasing field.
The outer domain walls marked in yellow typically do not form closed loops and are instead often connected to the sample edges. For example, eight stripes are connected to the upper edge of the sample in Fig.~\ref{FigExp2}\textbf{a}.
Our observations show that such stripes may give rise to an arbitrary number of skyrmions.
In the present case (Fig.~\ref{FigExp2}), these stripes disappear continuously (see Fig.~\ref{FigExp2}\textbf{g}) and the outer domain wall converges to form a single skyrmion.
This process can be followed based on the total number of antiskyrmions present at higher magnetic fields, as shown in Figs~\ref{FigExp2}\textbf{i} and \textbf{j}.
At intermediate fields, annihilation of skyrmion-antiskyrmion pairs is observed, as shown in Figs~\ref{FigExp2}\textbf{g} and \textbf{h}. 
The weaker stability of skyrmion-antiskyrmion pairs is in good agreement with micromagnetic simulations, as shown in Extended Data Fig.~\ref{FigS-Nucleation-theory}.
Furthermore, theoretical calculations suggest that the particle-antiparticle pair illustrated in Fig.~\ref{Fig1} is stable statically only above a critical film thickness, which we estimated for FeGe to be $40\pm5$~nm.
In a thinner film, such pairs annihilate immediately, similar to the behavior for the 2D model, as shown in Extended Data Figs~\ref{FigS-2D-1} and \ref{FigS-2D-2}.

Despite the topological equivalence between a skyrmion-antiskyrmion pair and a skyrmionium, we did not observe a transition between the two states experimentally.
However, micromagnetic simulations suggest that such a transition can be achieved by tilting the magnetic field slightly by several degrees.
In contrast to the formation of skyrmion-antiskyrmion pairs, the appearance of a skyrmionium in our experiments was a very rare event.
It should be noted that the approach described above for antiskyrmion nucleation is only applicable for thin plates, for which $t\lesssim L_\mathrm{D}$.
As reported in many earlier works, in thick plates of cubic chiral magnets, irrespective of their initial state, the system usually only converges to an energetically more favorable state that contains skyrmions.
An alternative approach for the creation of antiskyrmions in thick plates will be presented elsewhere.

In conclusion, we have observed the creation and annihilation of skyrmion antiparticles in an isotropic chiral magnet -- B20-type FeGe. 
Micromagnetic simulations support these observations and show excellent agreement with the experimental data. 
The experimental observation of skyrmion-antiskyrmion pairs in an FeGe plate whose thickness is below the size of a characteristic chiral modulation may serve as a platform to study the fundamental physics of topological solitons in magnetic solids.
The good qualitative agreement of the observed phenomena with theoretical predictions for a 2D model~\cite{Kuchkin_20i} suggests that a large diversity of other phenomena predicted by this model~\cite{Rybakov_19,Kuchkin_20ii,Kuchkin_21} may be verified experimentally in thin films of cubic chiral magnets.
Our results suggest that a thin plate of an isotropic chiral magnet may provide a platform for the experimental verification of the effect of a sign change in a topological Hall signal when the system contains antiskyrmions instead of skyrmions~\cite{Neubauer_09}.
Moreover, they open a wide vista for the experimental study of intriguing phenomena that manifest themselves as additional contributions to a Hall signal, even when the averaged topological density is zero (see, \emph{e.g.}, Refs~\cite{Bouaziz_2021, Pershoguba_2021}) for an identical number of particles and antiparticles. 

\renewcommand{\bibname}{Literary works}
%



\vspace{5mm}

\textbf{Acknowledgments.}
The authors thank Haifeng Du for help with sample preparation. This project has received funding from the European Research Council under the European Union's Horizon 2020 Research and Innovation Programme (Grant No.~856538 - project ``3D MAGiC''). 
F.N.R. was supported by Swedish Research Council Grants 642-2013-7837, 2016-06122, 2018-03659 and by the G\"{o}ran Gustafsson Foundation for Research in Natural Sciences. 
V.M.K. and N.S.K. acknowledge financial support from the Deutsche Forschungsgemeinschaft through SPP 2137 ``Skyrmionics" Grant No. KI 2078/1-1.
S.B.\ acknowledges financial support from the Deutsche Forschungsgemeinschaft through SPP 2137 ``Skyrmionics" Grant No.\ BL 444/16.
R.E.D-B. is grateful for financial support from the European Research Council under the European Union's Horizon 2020 Research and Innovation Programme (Grant No.~823717 - project ``ESTEEM3''; Grant No.~766970 - project ``Q-SORT'') and the Deutsche Forschungsgemeinschaft (Project-ID 405553726 – TRR~270). 

\textbf{Author contributions.}
F.Z. and N.S.K. conceived the project and designed the experiments.
F.Z. performed the TEM experiments and data analysis together with N.S.K and L.Y.
N.S.K., V.M.K. and F.N.R. developed the theory and performed numerical simulations. 
F.Z. and N.S.K. prepared the manuscript.
All of the authors discussed the results and contributed to the final manuscript.

\textbf{Data availability.}
All data are available from the corresponding authors upon reasonable request.

\textbf{Competing interests.}
The authors declare no competing interests.




\newpage

\section*{Methods}
\textbf{Micromagnetic calculations.}
The micromagnetic approach was followed in this work.
The total energy of the system includes the exchange energy, the DMI energy, the Zeeman energy and the self-energy of the demagnetizing field~\cite{Fratta}: 
\begin{align} 
\mathcal{E}\!=\!\int\limits_{V_\mathrm{m}}\!d\mathbf{r}\ 
&\mathcal{A}\sum\limits_{i=x,y,z} |\nabla m_i|^2 
+\mathcal{D}\,\mathbf{m}\!\cdot(\nabla\!\times\!\mathbf{m})
- M_\mathrm{s}\,\mathbf{m}\!\cdot\!\mathbf{B} + \nonumber\\
&+ \frac{1}{2\mu_0}\int\limits_{\mathbb{R}^3}\!d\mathbf{r}\ 
\sum\limits_{i=x,y,z}|\nabla A_{\mathrm{d}, i}|^2~,
\label{Ham_m}
\end{align}
where the magnetic field
\begin{align}
\mathbf{B} = \mathbf{B}_\text{ext} + \nabla\!\times\!\mathbf{A}_\text{d}~,
\end{align}
${\mathbf{m}(\mathbf{r})}=\mathbf{M}(\mathbf{r})/M_\text{s}$ is a unit vector field that defines the direction of the magnetization,  $M_\text{s}=|\mathbf{M}(\mathbf{r})|$ is the saturation magnetization,
${\mathbf{A}_\text{d}(\mathbf{r})}$ is the component of magnetic vector potential induced by the magnetization,
$\mathcal{A}$ is the exchange stiffness constant, $\mathcal{D}$ is the constant of isotropic bulk DMI and $\mu_0$ is the vacuum permeability (${\mu_0 \approx 1.256}$~$\mu$NA$^{-2}$).
In our simulations, we used the following material parameters for FeGe~\cite{Zheng_18}:
$\mathcal{A}=4.75$~pJm$^{-1}$, $\mathcal{D}=0.853$~mJm$^{-2}$ and $M_\text{s}=384$~kAm$^{-1}$. 
The solutions of the Hamiltonian~(\ref{Ham_m}) were found by the numerical energy minimization method described in Ref.~\cite{Zheng_21} using Excalibur software~\cite{Excalibur}.

\vspace{5mm}


\textbf{Initial guesses for calculating antiparticles.}
Defining the angle $\phi_\mathrm{A}=\pi z/L_\mathrm{D}$, the orientation of an antiskyrmion is first set in every $z$-section in the form: 
\begin{equation}
    \left(\begin{array}{c}
    x^{\prime}\\
    y^{\prime}
    \end{array}\right)=\dfrac{1}{l}\left(\begin{array}{cc}
    \cos\phi_\mathrm{A} & \sin\phi_\mathrm{A}\\
    -\sin\phi_\mathrm{A} & \cos\phi_\mathrm{A}
    \end{array}\right)\left(\begin{array}{c}
    x\\
    y
\end{array}\right).
\end{equation}
Following the approach introduced in Ref.~\cite{Barton-Singer_20}, the auxiliary vector field is defined according to the expression
\begin{equation}
    \boldsymbol{m}^{\prime}=\dfrac{1}{g_{+}}\left(2x^{\prime}-y^{\prime}, x^{\prime}-2y^{\prime},g_{-}\right)^\mathrm{T},   
\end{equation}
where $g_\pm=\dfrac{5}{4}\left(\left(x^{\prime}\right)^{2}+\left(y^{\prime}\right)^{2}\right)-2x^{\prime}y^{\prime}\pm1$ and the scaling parameter $l$ defines the antiskyrmion size. In our simulations, we let $l=0.25 L_\mathrm{D}$. For an antiskyrmion embedded in a ferromagnetic background, we use the following initial guess: 
\begin{equation}
    \boldsymbol{m}=R_\mathrm{z}(\phi_\mathrm{A})\boldsymbol{m}^{\prime}~,  
    \label{ASK_FM}
\end{equation}
where $R_\mathrm{z}(\phi_\mathrm{A})$ is a $3\times 3$ rotational matrix  about the $z$~axis. 
For an antiskyrmion embedded in the conical phase, the initial guess takes the form
\begin{equation}
  \boldsymbol{m}=R_\mathrm{z}(\phi_\mathrm{c})R_\mathrm{y}(\theta_\mathrm{c})R_\mathrm{z}(\phi_\mathrm{A}-\phi_\mathrm{c})\boldsymbol{m}^{\prime}~, 
\end{equation}
where $\theta_\mathrm{c}$ is the cone phase angle and $\phi_\mathrm{c}=2\pi z/L_\mathrm{D}$.

An alternative approach for the construction of the initial state for an antiskyrmion is illustrated in Extended Data Fig.~\ref{FigS-ini}.
This approach has been verified in Mumax~\cite{mumax} software.

\vspace{5mm}

\textbf{Magnetic imaging in the TEM.}
Fresnel defocused Lorentz TEM imaging and off-axis electron holography was performed in an FEI Titan 60-300 TEM operated at 300~kV. The microscope was operated in aberration-corrected Lorentz mode with the sample in magnetic-field-free conditions. The conventional microscope objective lens was then used to apply out-of-plane magnetic fields to the sample of between -0.15 and +1.5~T (pre-calibrated using a Hall probe). A liquid-nitrogen-cooled specimen holder (Gatan model 636) was used to control the specimen temperature between 95 and 380~K. Images were recorded when the specimen temperature was 95~K, if not otherwise specified. Fresnel defocus Lorentz TEM images and off-axis electron holograms were recorded using a 4k~$\times$~4k Gatan K2 IS direct electron counting detector. The defocus distance was $|\Delta{z}|=800$~$\mu$m for all images presented in the text, if not otherwise specified.
Multiple off-axis electron holograms, each with a 6~s exposure time, were recorded to improve the signal-to-noise ratio and analyzed using a standard fast Fourier transform algorithm in Holoworks software (Gatan). 


\vspace{5mm}


\renewcommand{\bibname}{suppl references}


\setcounter{figure}{0}
\captionsetup[figure]{labelfont={bf},name={Extended Data Fig.},labelsep=period}

\begin{figure*}[ht] 
    \centering
    \includegraphics[width=18cm]{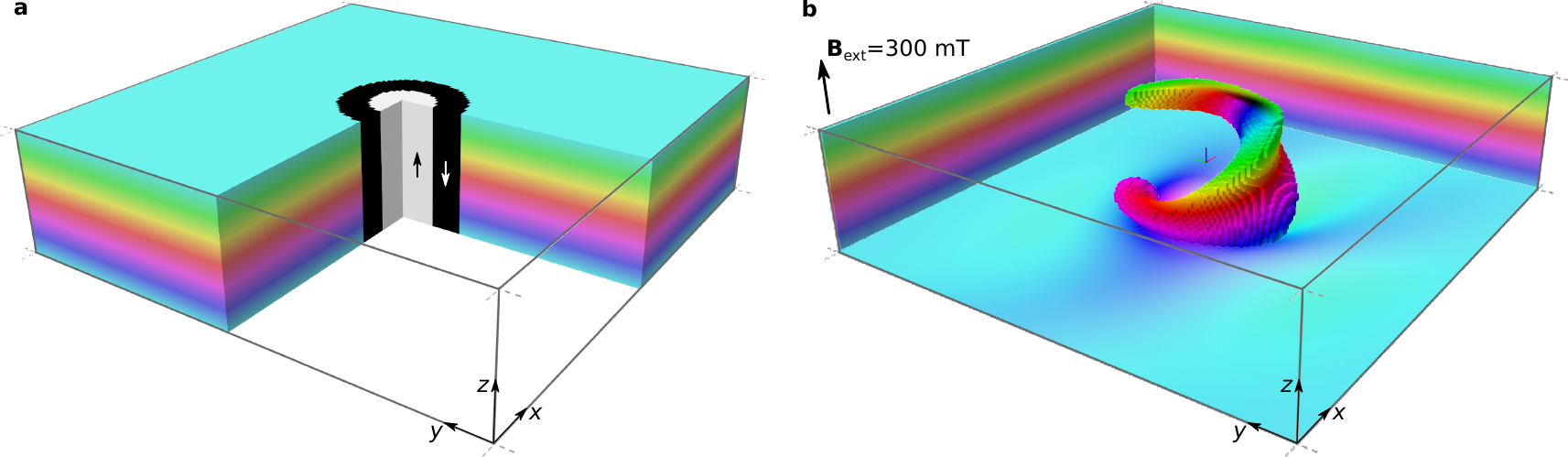}
    \caption{\small \textbf{Initial guess for antiskyrmion in micromagnetic simulations.} \textbf{a}, Initial state comprising two coaxial cylindrical domains with their magnetization up (white) and down (black), embedded in a conical state whose $k$~vector is $2\pi/L_\mathrm{D}$ and points along the $z$~axis. The cone angle $\theta_\mathrm{c}=\mathrm{acos}[B_\mathrm{ext}/(B_\mathrm{D}+\mu_0 M_\mathrm{s})]$. \textbf{b}, Antiskyrmion in an external field of $B_\mathrm{ext}=0.45(B_\mathrm{D}+\mu_0 M_\mathrm{s})\approx300$mT obtained after full energy minimization in Mumax assuming periodic boundary conditions in the $xy$~plane. The 3D texture of the antiskyrmion is visualized, showing cuboids at the edges of the simulated domain and cuboids where $m_\mathrm{z}<0$. The size of the domain in the $xy$~plane is $4L_\mathrm{D}\times 4L_\mathrm{D}$ and the thickness is $1L_\mathrm{D}$. For other parameters, see the Supplementary Material.  
    }
    \label{FigS-ini}
\end{figure*}

\begin{figure*}[ht] 
    \centering
    \includegraphics[width=18cm]{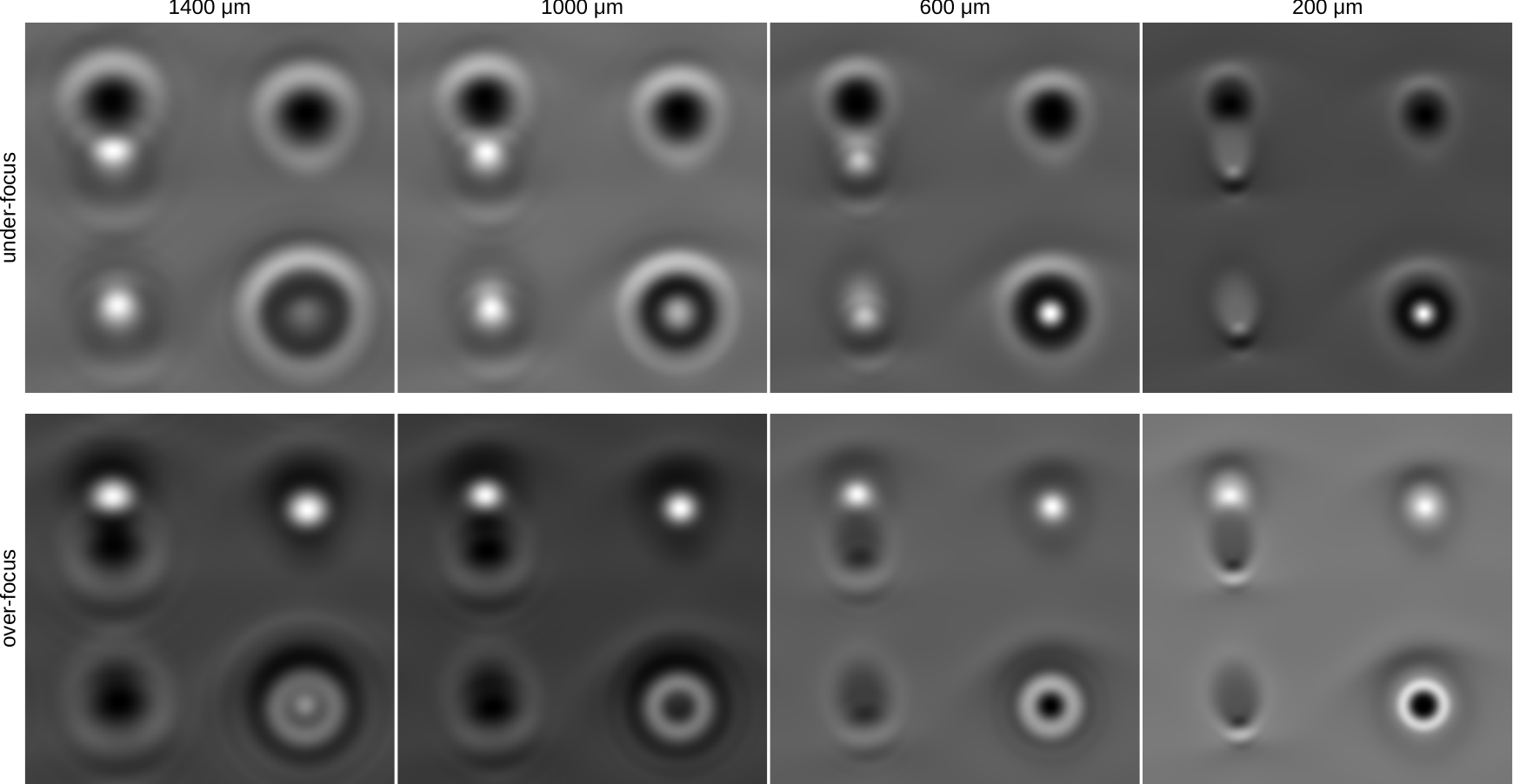}
    \caption{\small \textbf{Simulated Lorentz TEM images for a skyrmion, an antiskyrmion, a skyrmion-antiskyrmion pair and a skyrmionium for different  defocus distances.} For the corresponding spin texture, see Fig. \ref{Fig1}a in the main text.
    }
    \label{FigS-LTEM-defocus}
\end{figure*}

\begin{figure*}[ht] 
    \centering
    \includegraphics[width=18cm]{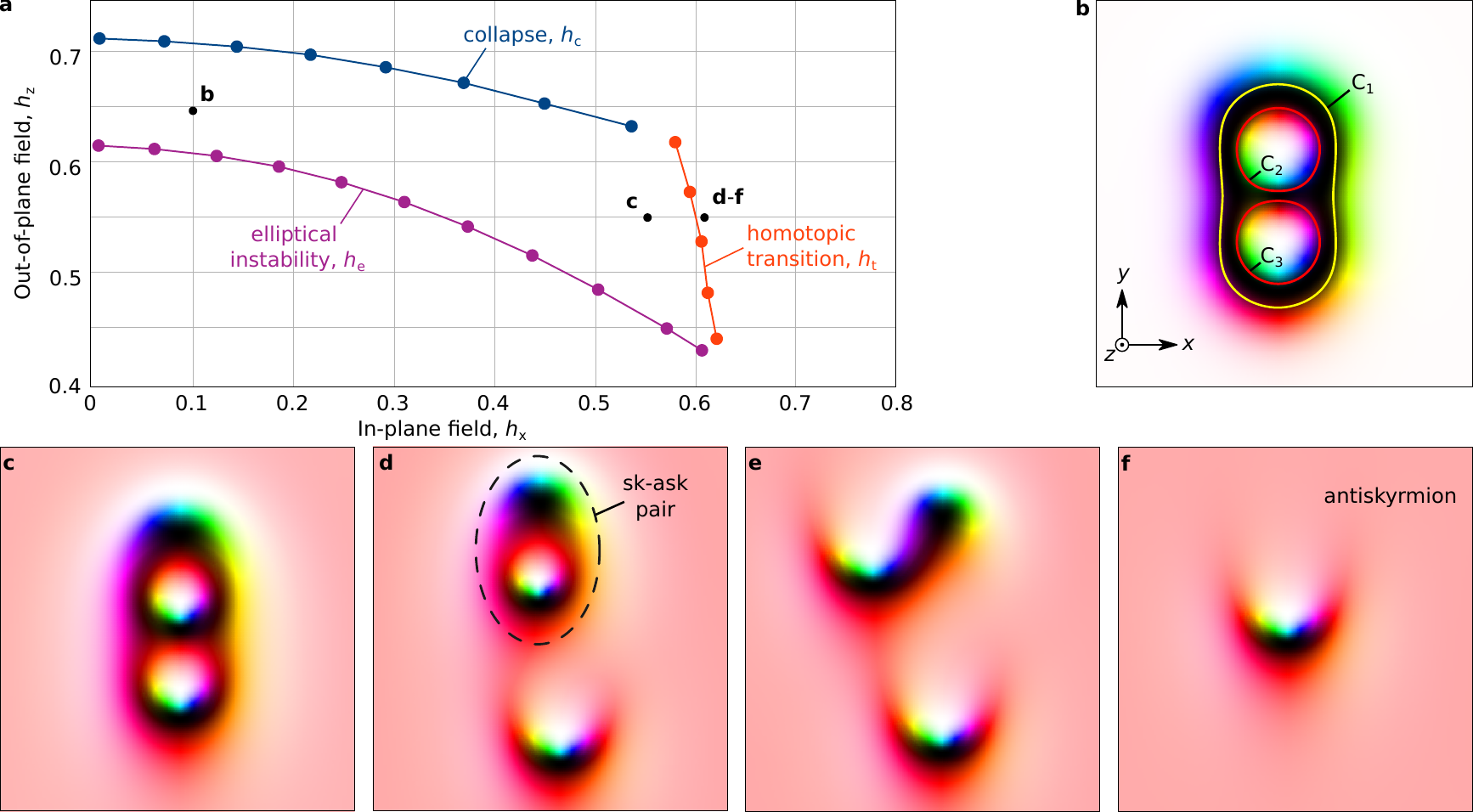}
    \caption{\small \textbf{Homotopical transition of a 2D skyrmion bag with $Q=1$.}
    \textbf{a}, Stability diagram for the 2D skyrmion bag depicted in \textbf{b}, given in terms of the in-plane and out-of-plane components of the reduced external magnetic field $\mathbf{h}=\mathbf{B}_\mathrm{ext}/B_\mathrm{D}$. The calculations were performed for a large domain of size $8L_\mathrm{D}\times8L_\mathrm{D}$ for the model Hamiltonian given in Ref.~\cite{Kuchkin_20i}.
    The stability region of the skyrmion bag is bounded by the collapse field from above, the elliptic instability field from below and the instability with respect to a homotopical (continuous) transition into another state with conservation of the topological index $Q$ on the right.
    The spin textures depicted in \textbf{b} and \textbf{c} correspond to stable configurations in different fields (see corresponding labels in \textbf{a}).
    Contour lines C$_1$-C$_3$ indicate closed 180$^\circ$ domain walls. 
    The yellow contour encloses the area with $m_\mathrm{z}<0$ (black), while the red contours enclose the area with $m_\mathrm{z}>0$ (white) -- as in Fig.~\ref{FigExp1} in the main text.
    Images \textbf{d}-\textbf{f} illustrate the homotopical transition that the skyrmion bag in \textbf{c} undergoes as soon as the external magnetic field exceeds the critical field $h_\mathrm{t}$ (see the red transition line in \textbf{a}).
    Images \textbf{d}-\textbf{e} are unstable configurations, representing snapshots taken at different stages during direct energy minimization, while the antiskyrmion in \textbf{f} is a stable configuration.
    }
    \label{FigS-2D-1}
\end{figure*}

\begin{figure*}[ht] 
    \centering
    \includegraphics[width=18.cm]{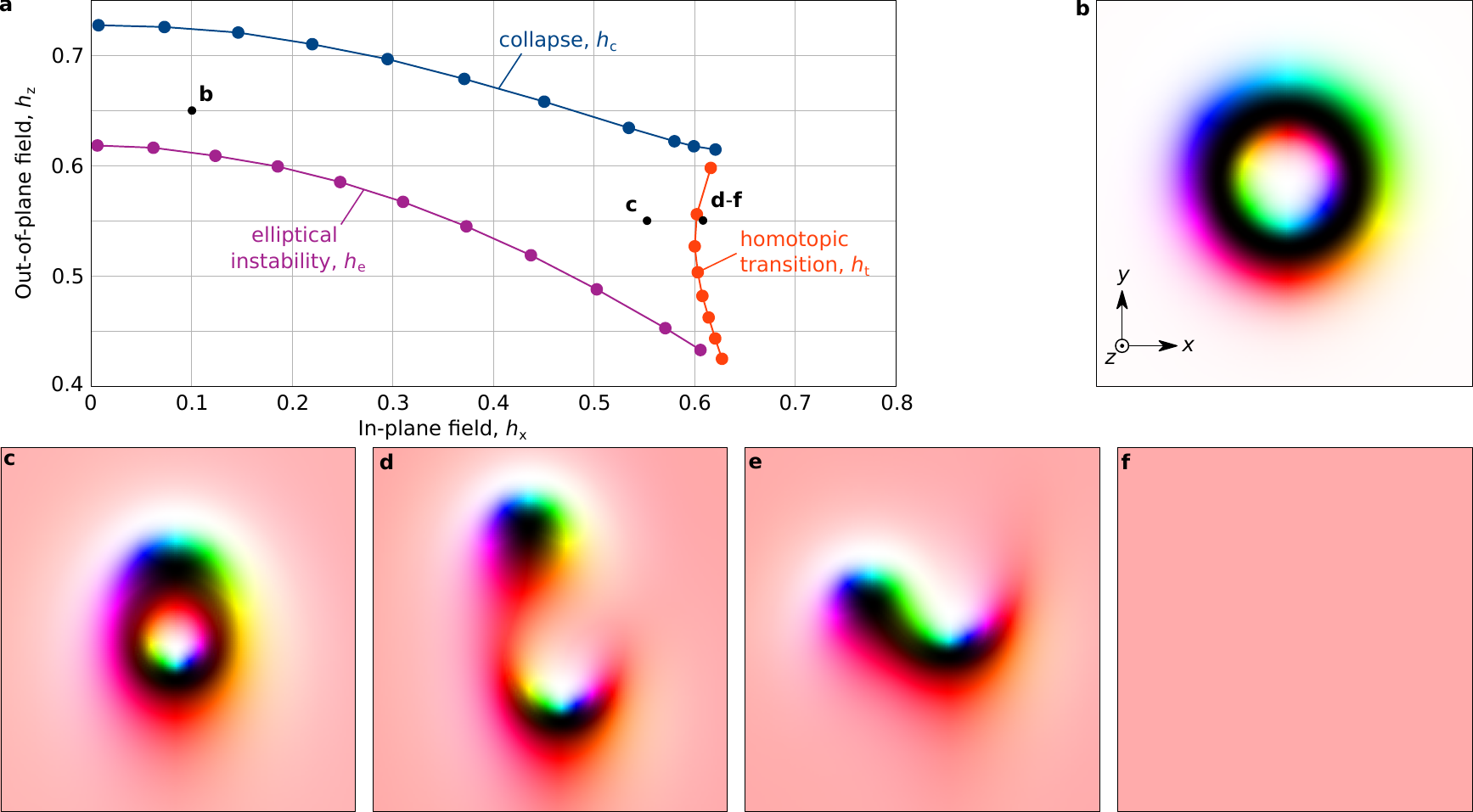}
    \caption{\small \textbf{Homotopical transition of a 2D skyrmionium with $Q=0$.}
    As for \textbf{Extended Data Fig.}~\ref{FigS-2D-1}, but for the skyrmionium depicted in \textbf{b}, which continuously converges to the tilted ferromagnetic state (\textbf{f}) under a homotopical transition \emph{via} the formation of a skyrmion-antiskyrmion pair (\textbf{d}), followed by its annihilation (\textbf{e}).   
    }
    \label{FigS-2D-2}
\end{figure*}

\begin{figure*}[ht] 
    \centering
    \includegraphics[width=18cm]{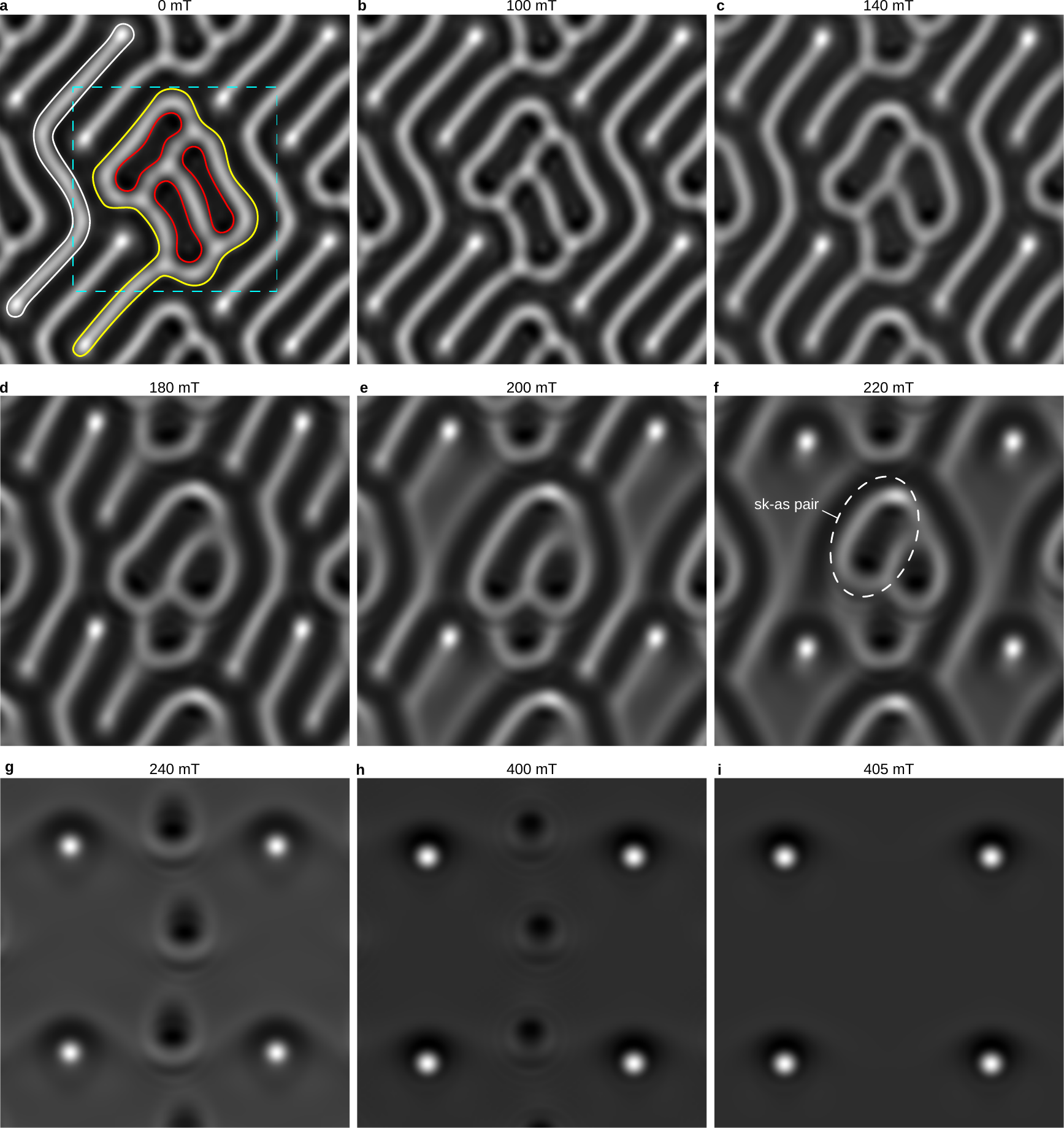}
    \caption{\small \textbf{Nucleation of antiskyrmions and skyrmion-antiskyrmion pairs in micromagnetic simulations.}
    Over-focus Lorentz TEM images were calculated for equilibrium magnetic textures relaxed at different values of increasing external magnetic field for a sample of thickness 50~nm, periodic boundary conditions and a defocus distance of 600 $\mu$m. The actual size of the simulated domain is marked by a dashed blue line in \textbf{a}. For illustrative purposes, a larger field of view is shown. The white, yellow and red contour lines have the same meaning as in the experimental images shown in Fig.~\ref{FigExp2} in the main text and in \textbf{Extended Data Figs}~\ref{FigS-nucleation-1}, \ref{FigS-nucleation-2} and \ref{FigS-nucleation-3}. As the field increases, we observe nucleation of skyrmion-antiskyrmion pairs (\textbf{f}), which quickly annihilate with further increasing external field (\textbf{g}).
    For a sample of thickness 50~nm, antiskyrmions collapse for $B_\mathrm{ext}>400$ mT.
    }
    \label{FigS-Nucleation-theory}
\end{figure*}

\begin{figure*}[ht] 
    \centering
    \includegraphics[width=18cm]{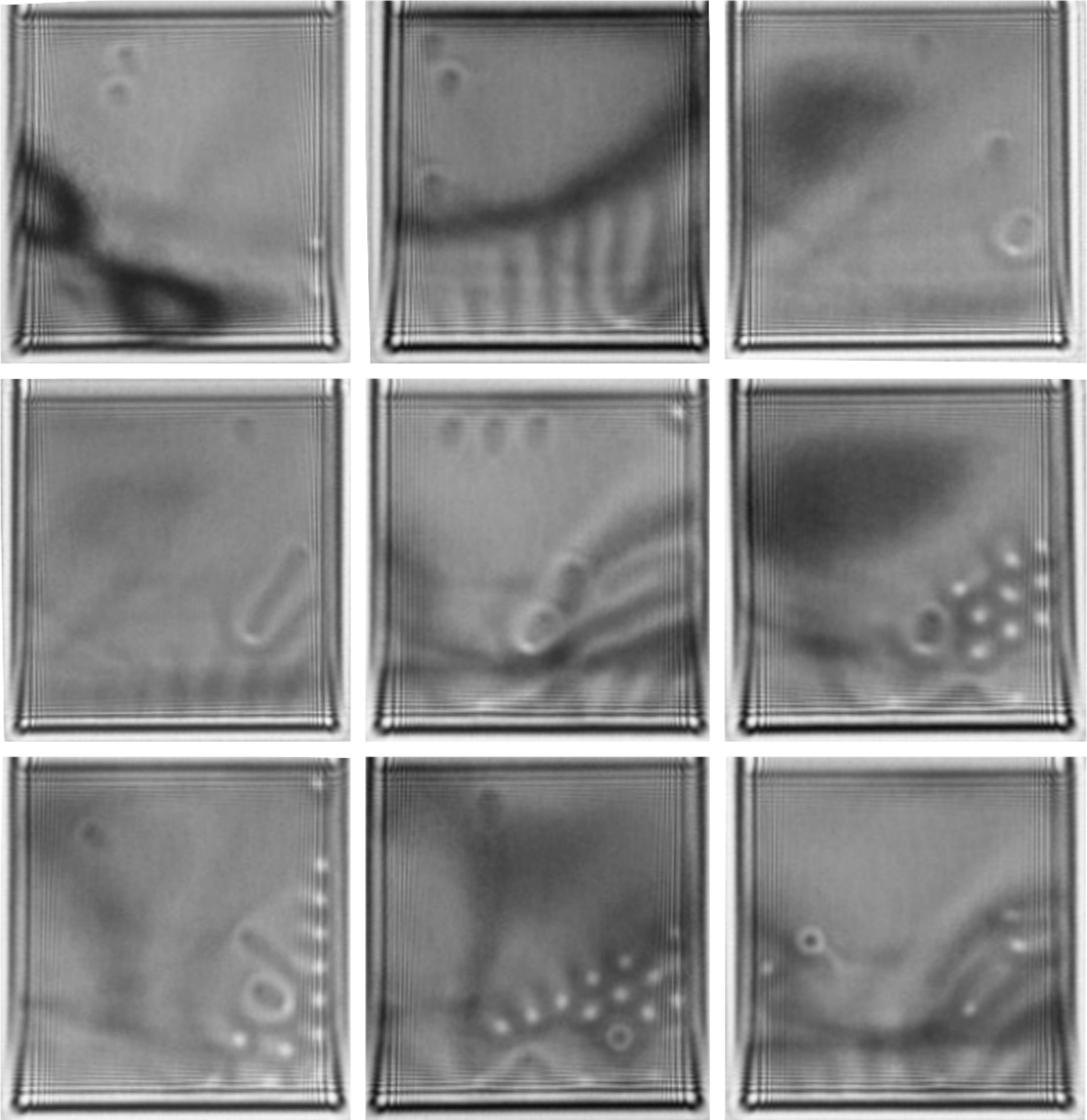}
    \caption{\small \textbf{Lorentz TEM images showing contrast features that correspond to a skyrmion, an antiskyrmion, a skyrmion-antiskyrmion pair and a skyrmionium.} The images were recorded over-focus using a defocus distance of 800~$\mu$m at a specimen temperature of 95~K and in applied fields of between 200 and 240~mT.
    }
    \label{FigS-collection}
\end{figure*}

\begin{figure*}[ht] 
    \centering
    \includegraphics[width=15cm]{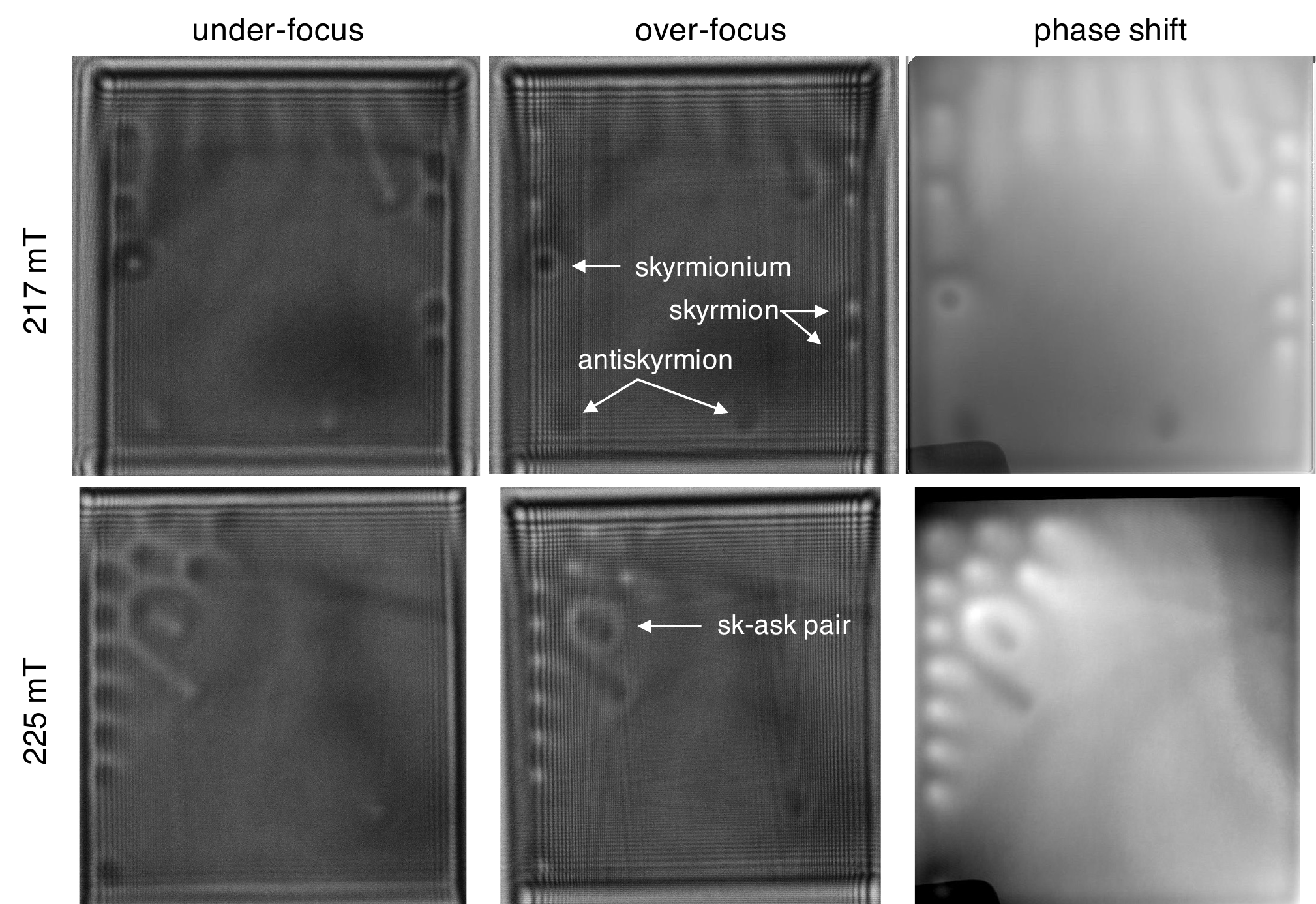}
    \caption{\small \textbf{Representative Lorentz TEM images showing contrast features that correspond to a skyrmion, an antiskyrmion, a skyrmion-antiskyrmion pair and a skyrmionium.}
    The defocus distance is 800 $\mu$m. The images were recorded at a specimen temperature of 95~K.
    }
    \label{FigS-TEM}
\end{figure*}

\begin{figure*}[ht] 
    \centering
    \includegraphics[width=18cm]{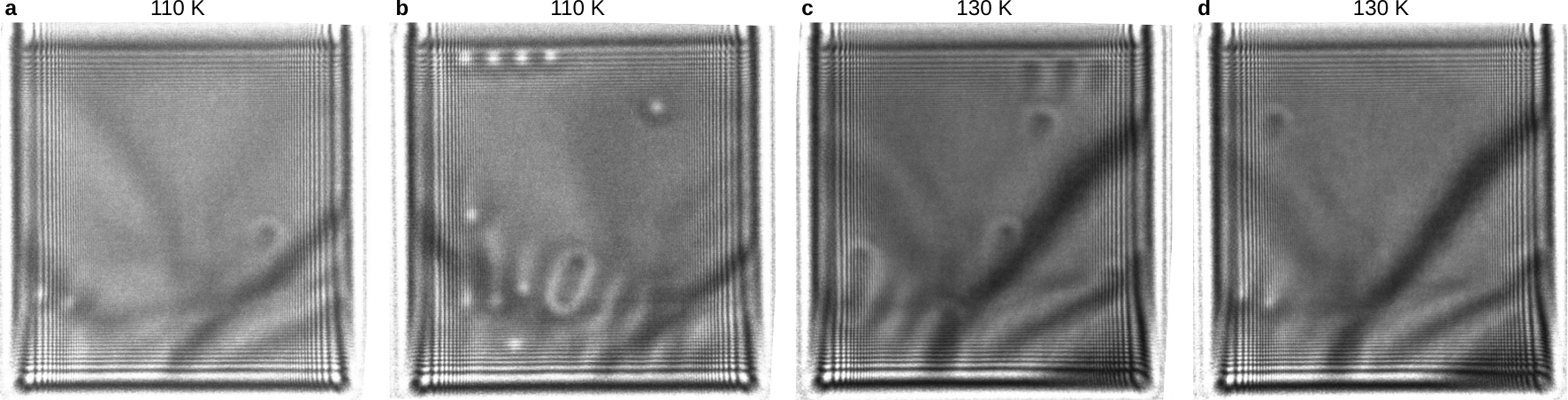}
    \caption{\small \textbf{Representative Lorentz TEM images showing contrast features that correspond to magnetic antiskyrmions and skyrmion-antiskyrmion pairs at different temperatures.} The images were recorded over-focus. The specimen temperature is denoted above each panel. The external magnetic field was $\sim 160$~mT.
    }
    \label{FigS-temperature}
\end{figure*}

\begin{figure*}[ht] 
    \centering
    \includegraphics[width=18cm]{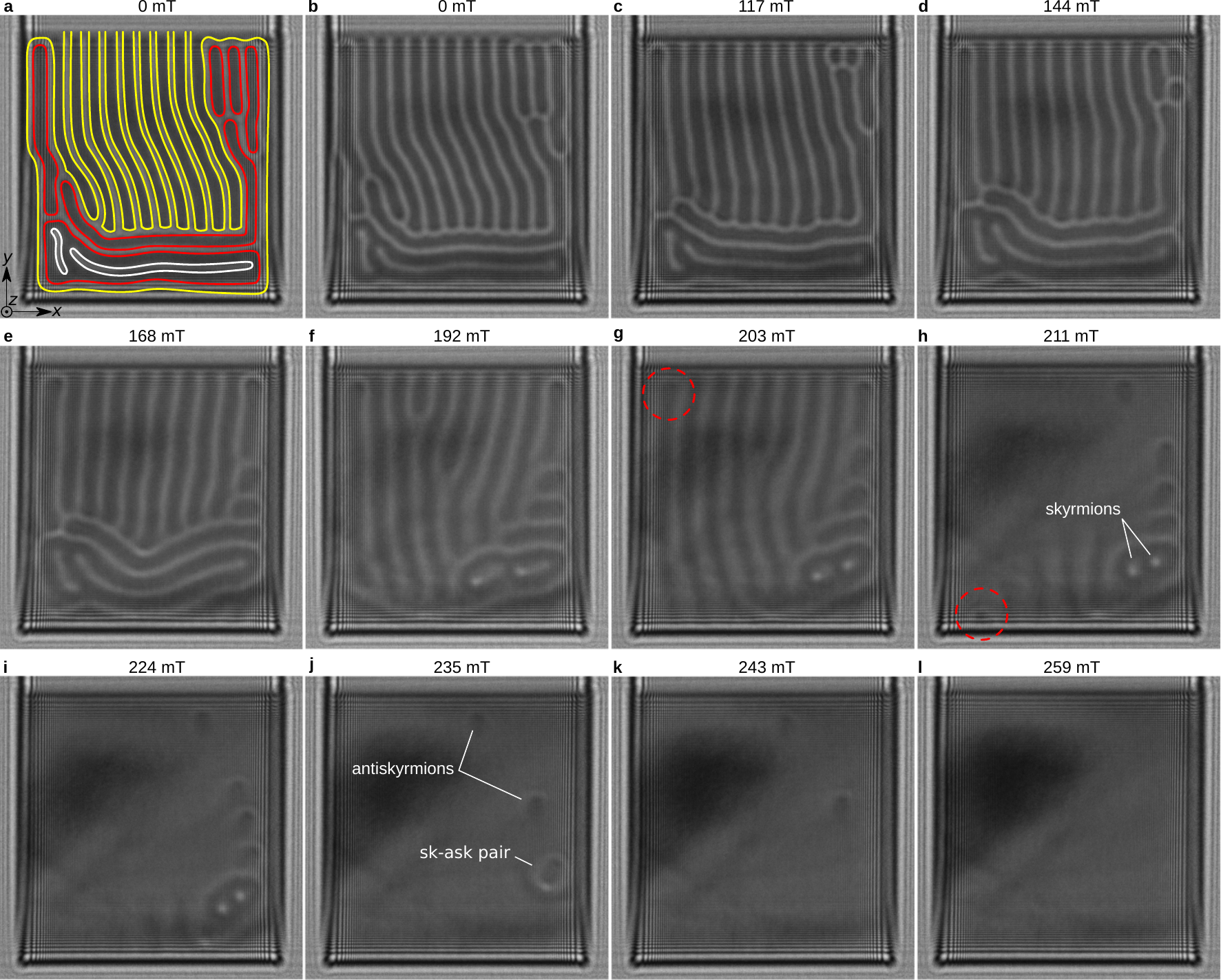}
    \caption{\small \textbf{Lorentz TEM images showing the nucleation of magnetic antiskyrmions.} 
    The initial state of the system was different from that presented in Fig.~\ref{FigExp2}\textbf{a}.
    The images were recorded over-focus in increasing perpendicular magnetic fields. The magnitude of the field is each indicated above each image.
    \textbf{a} and \textbf{b} are identical. Lines in \textbf{a} mark domain walls, similar to those in Fig.~\ref{FigExp2}a.
    Red circles mark antiskyrmions, which collapse at 224~mT (see \textbf{h} and \textbf{i}).
    The two skyrmions labeled in \textbf{h} annihilate with two antiskyrmions with increasing field (see \textbf{j}).
    \textbf{j} shows a skyrmion-antiskyrmion pair, which annihilates with increasing field at 243~mT (see \textbf{k}).
    The antiskyrmions collapse at a field of 259~mT (see \textbf{l}).
    The defocus distance is 600~$\mu$m.
    %
    }
    \label{FigS-nucleation-1}
\end{figure*}

\begin{figure*}[ht] 
    \centering
    \includegraphics[width=18cm]{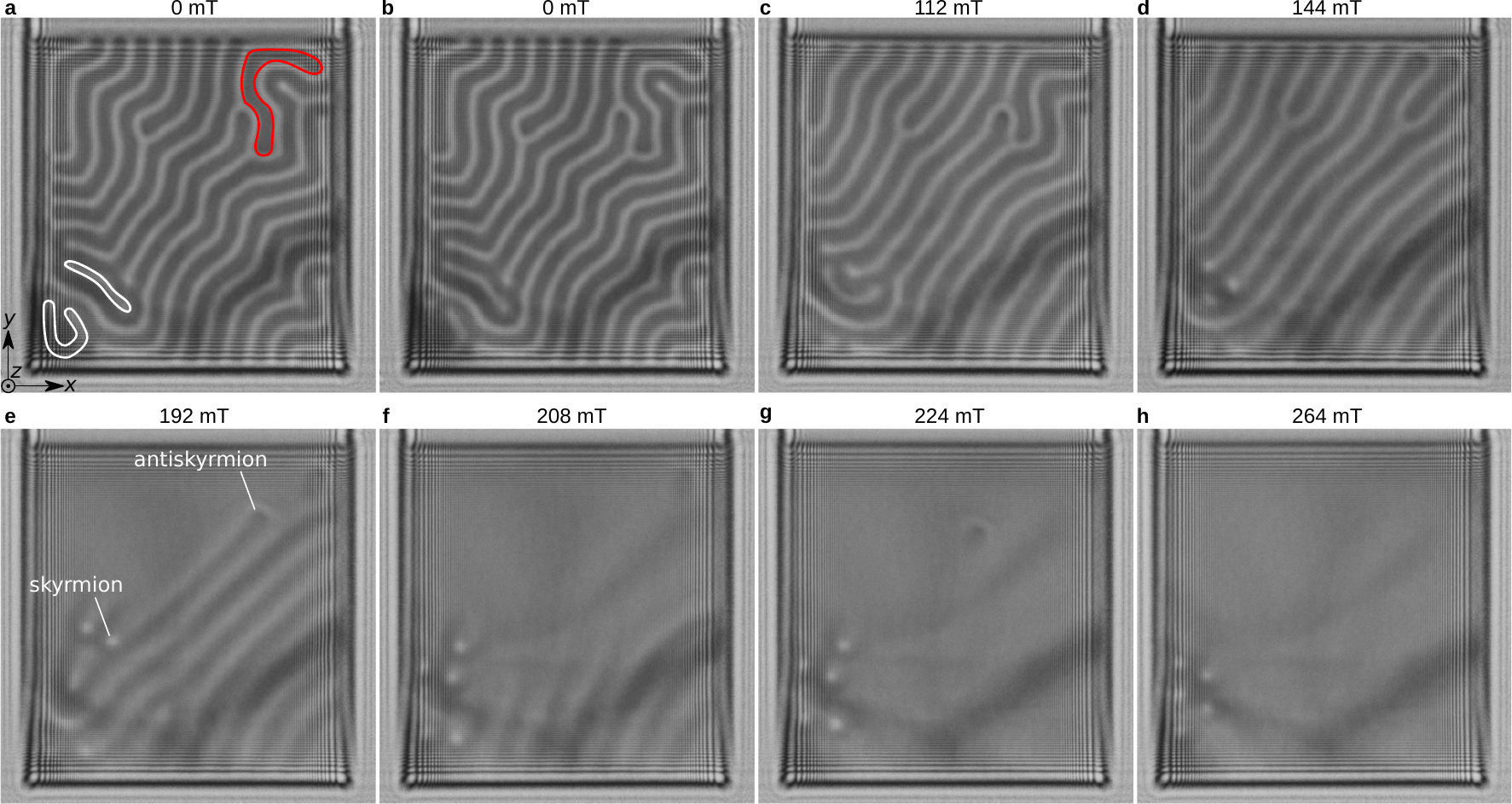}
    \caption{\small \textbf{Lorentz TEM images showing the nucleation of magnetic antiskyrmions.} 
    %
    %
    %
    The initial state of the system has one closed domain wall, which is marked by a red contour and converges to a single antiskyrmion with increasing field (see \textbf{f}). 
    The two white contours converge to two skyrmions with increasing field (see \textbf{f}).
    The skyrmion and antiskyrmion marked in \textbf{e} annihilate with each other when the field is increased to 208~mT (see \textbf{f}).
    %
    }
    \label{FigS-nucleation-2}
\end{figure*}

\begin{figure*}[ht] 
    \centering
    \includegraphics[width=18cm]{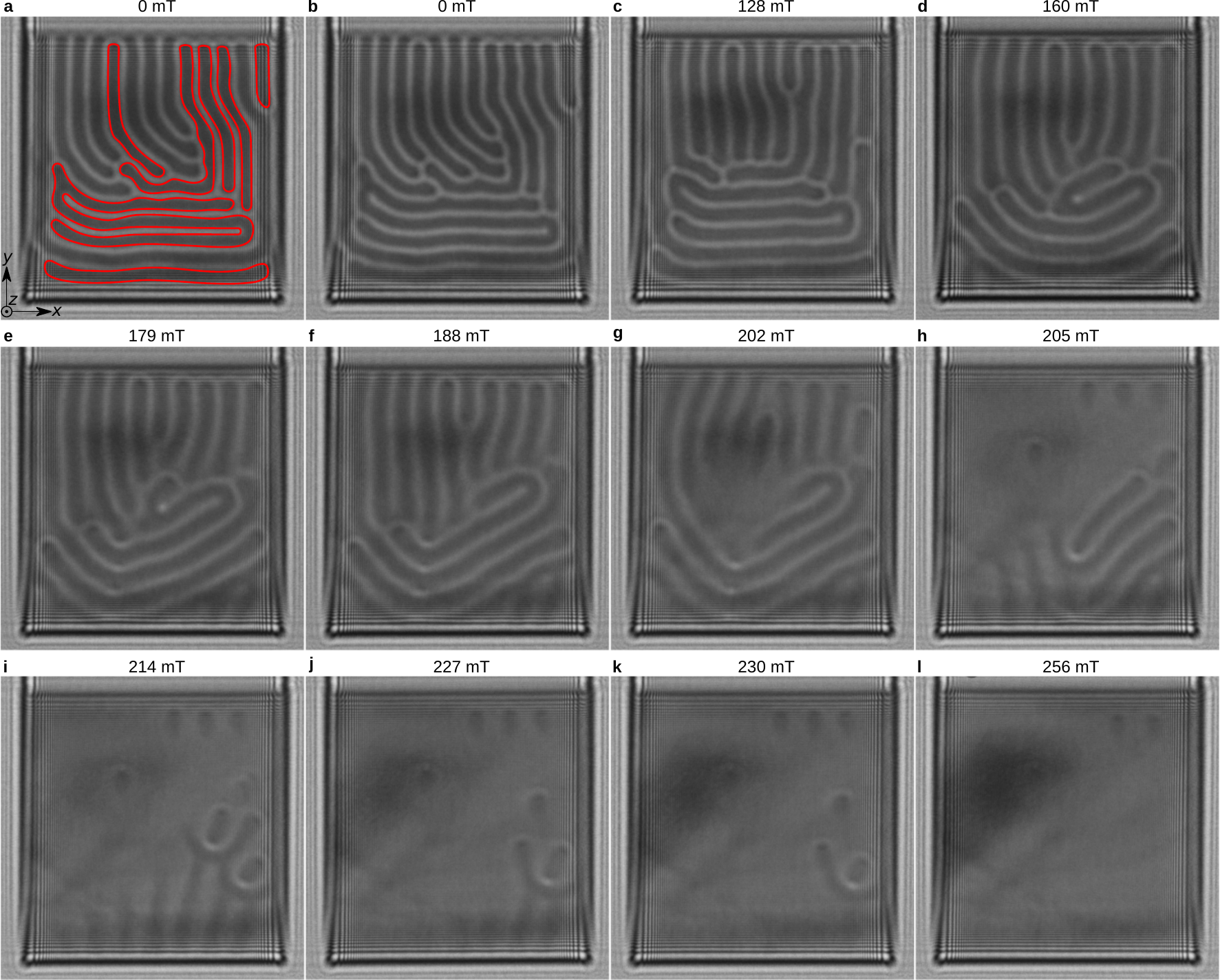}
    \caption{\small \textbf{Lorentz TEM images showing the nucleation of magnetic antiskyrmions.} 
    %
    %
    %
    The initial state of the system has no closed domain walls that converge to form skyrmions with increasing field.
    \textbf{h}-\textbf{k} show skyrmion-antiskyrmion pairs, which annihilate with each other with increasing field (compare \textbf{i} and \textbf{j}, \textbf{k} and \textbf{l}).
    %
    %
    }
    \label{FigS-nucleation-3}
\end{figure*}

\end{document}